\documentclass{PoS}

\title{Constraints on the Galactic Halo Dark Matter from Fermi-LAT Diffuse Measurements}

\ShortTitle{Galactic Halo Dark Matter and the Fermi-LAT Diffuse Measurement}

\author{\speaker{Gabrijela Zaharijas} \\
       Stockholm University, IPhT/CEA Saclay\\
        E-mail: \email{gabrijela.zaharijas@cea.fr}}

\author{Alessandro Cuoco, Zhaoyu Yang, Jan Conrad  \\
       Stockholm University,\\
        E-mail: \email{cuoco@phys.au.dk,conrad@fysik.su.se,zhaoyu@fysik.su.se}}

\author{on behalf of the Fermi-LAT collaboration}

\abstract{

The diffuse gamma ray emission from astrophysical backgrounds in our Galaxy and the signal due to the annihilation or decay of Dark Matter (DM) in the Galactic Halo are expected to have a substantially different morphology and spectral signatures. In order to exploit this feature we perform a full sky and spectral binned likelihood  fit of both components, using  data collected during the first 21 months of operation of the Fermi-LAT observatory. Preliminary constraints  are presented  on the DM annihilation cross section and decaying rate for various masses and annihilation/decay modes.}

\FullConference{Identification of Dark Matter 2010-IDM2010\\
		July 26-30, 2010\\
		Montpellier France}

\begin{document}

\section{Introduction}

The Dark Matter (DM) properties, despite of being one of the most widely investigated topics in contemporary fundamental physics, are still largely undetermined. However, the current and near future experiments are cutting more and more into the parameter space region predicted for the most popular type of DM candidates, i.e. Weakly Interacting Massive Particles (WIMPs) (for a review see \cite{Bertone:2004pz}). In particular, high energy gamma ray astronomy is an exciting probe into these fundamental questions. In fact, gamma rays are products of hadronization and radiative loss processes, and are therefore unavoidably emitted in DM annihilation and decay. 
Also, the propagation of gamma rays is unaffected by the Galaxy, and therefore the data contain the information on the morphology of the emission region, together with the spectra (unlike, i.e. cosmic rays). Not least, the Large Area Telescope (LAT), onboard the Fermi gamma-ray observatory \cite{Atwood:2009ez}, is now providing  unprecedented high quality data and statistics which makes it timely to investigate DM in gamma-rays.

We will focus in particular on DM signatures in the gamma-ray diffuse emission. Diffuse data contain about $90\%$ of the LAT photons, and is extremely rich with information. Its {\em Galactic component} encodes information on the propagation and origin of cosmic rays, distribution of cosmic ray sources, magnetic and starlight fields in our Galaxy, while its {\em extragalactic component} is a superposition of all unresolved sources emitting gamma rays in the Universe, providing a signature of energetic phenomena over cosmological scales. {\em Both} components are expected to receive a contribution from {\em dark matter} annihilation/decay: the  Galactic signal from the smooth dark matter halo and Galactic substructures, while the extragalactic one from self-annihilation of dark matter in all halos at all redshifts. 
The dark matter searches with the diffuse data by the Fermi-LAT team is  being pursued with the Extragalactic (Isotropic) Signal, by using the intensity and spectral shape of the signal \cite{Abdo:2010dk} and angular anisotropies \cite{jsg}. Here we will investigate the DM signal originating in our Halo (see also the analysis presented in \cite{ba}).

Due to the bright sources present in the Galactic Center and the bright diffuse emission along the plane, it has been argued e.g. in \cite{Serpico:2008ga}, that the region of the inner Galaxy, extending $10-20~\deg$ away from the Galactic Plane, is the most promising in terms of a signal to background ratio $S/N$. 
As an additional advantage, the perspectives to constrain a DM signal in that region, become less sensitive to the unknown profile of the DM halo. In particular the $S/N$ of quasi-cored profiles is only a factor $\sim 2$ worse than for the NFW case (compared to an order of magnitude of uncertainty, when one considers the Galactic center region). The increase in $S/N$ ratio away from the plane is further emphasized for DM models in which DM annihilations result in a significant fraction of leptons in the final state. These leptons diffuse in the Galaxy and produce high energy gamma rays mainly through the Inverse Compton (IC) scattering on the Interstellar Radiation Field (ISRF). By diffusing away from the Galactic Center region, electrons produce an extended gamma ray signal which is easier to distinguish from the astrophysical signal (\cite{Borriello:2009fa}).

In this work we build on this idea, and consider the whole sky data, while masking out the Galactic Plane. We test DM annihilation and decay signal in the smooth Galactic halo, by performing a spatial and energy fit. We also take into account the most up to date modeling of the diffuse signal of astrophysical origin \cite{paper2}. 
However, at present, this approach faces few limitations. The high statistics of the Fermi data is currently unmatched with our knowledge of the cosmic ray production and propagation mechanisms, as determined by the cosmic ray experiments and cosmic ray source observations. In other words, current diffuse models, which rely on the gamma ray independent measurements, have both significant energy and spatial residuals, on small and large scales. This poses a significant challenge to the statistical approach to constrain the presence of an additional (e.g. dark matter) component in the fit (see also \cite{ba}). In this talk, we present preliminary limits, based on the astrophysical background model, found to be one of the most conservative in terms of dark matter searches, among the models currently considered in analysis of the diffuse emission, by the Fermi-LAT collaboration \cite{paper2}.

  \section{Modeling of the high energy galactic diffuse emission}

The Galactic diffuse gamma ray emission is produced by the interaction of energetic cosmic-ray electrons and protons with interstellar nucleons and photons. Its main components are the decay of neutral pions produced in CR proton interactions with the interstellar gas, bremsstrahlung of the CR electron population on the interstellar gas and their inverse Compton scattering of the interstellar radiation field.  The three most important Interstellar Gas components are atomic (HI, making up $\sim 50 \%$ of the diffuse emission), molecular (H$_2$, about $10\%$) and ionized (HII$\sim 2 \%$) hydrogen.   Starting from radio observations, gas maps can be built  for different annuli in galacto-centric distance  \cite{paper2} providing, effectively, a 3D model of the gas distribution in the Galaxy. An additional uncertainty is introduced by the conversion of the observed quantities to the actual hydrogen column densities, which are taken into account in the way the fitting procedure is performed. Finally, a full 3D model of ISRF is used, computed from a modeling of radiation emission from stars and further reprocessing in the galactic dust \cite{porter}.

Based on this set of inputs (scattering targets),  the gas and IC gamma-ray maps are produced by the \texttt{GALPROP} code \cite{galprop}, using up to date knowledge of cosmic ray distribution sources and propagation parameters. Main parameters entering the \texttt{GALPROP} modeling (in addition to the above mentioned) are: the distribution of cosmic ray sources, the height of the diffusive halo, the nucleon and electron injection spectrum, the diffusion coefficient and Alfv{\'e}n speed (for a review see \cite{galprop}). In this work, we closely follow the actual procedure for studies of gamma ray diffuse emission of the Fermi-LAT collaboration \cite{paper2}. For a particular choice of the CR source distribution and the diffusive halo size $z_h$, other propagation parameters are obtained with a fit to CR data only. The models are then compared to the gamma ray data using a  binned likelihood fit using the \texttt{GARDIAN} tool. Together with the above \texttt{GALPROP} models, describing the Galactic diffuse emission, point sources, modeled from the 1FGL point source catalog \cite{1FGL}, as well as an isotropic emission component \cite{Abdo:2010nz} have been included in the fit with a free normalization.

DM maps  are fitted jointly with the diffuse components described above. The ingredients entering the production of dark matter maps are the DM distribution and its annihilation/decay spectra. Numerical simulations of Milky Way size halos reveal a smooth halo which contain large number of subhalos \cite{Diemand:2007qr}. The properties of the smooth halo seem to be well understood, at least on the scales resolved by simulations (outside of inner few hundreds of parsecs, e.g. few degrees away of the Galactic Center), and to converge among the various simulations. The properties of the subhalo population, on the other hand, are more model dependent, and we conservatively consider only the smooth component in this work.  We parametrize the smooth DM component with a  NWF profile \cite{Navarro:1995iw} and a cored isothermal sphere profile. For the local density of dark matter we take the value of $\rho_0=0.43$ GeV/cm$^3$, and the scale radius is assumed to be 20 kpc (NFW) and 2.8 kpc (isothermal profile). The actual choice of the DM density profile does not have a huge effect on our limits as we do not consider the central few degrees of the Galaxy (where these distributions differ the most).

\begin{figure}[t]
\begin{center}$
\begin{array}{cc}
\includegraphics[width=0.45\columnwidth]{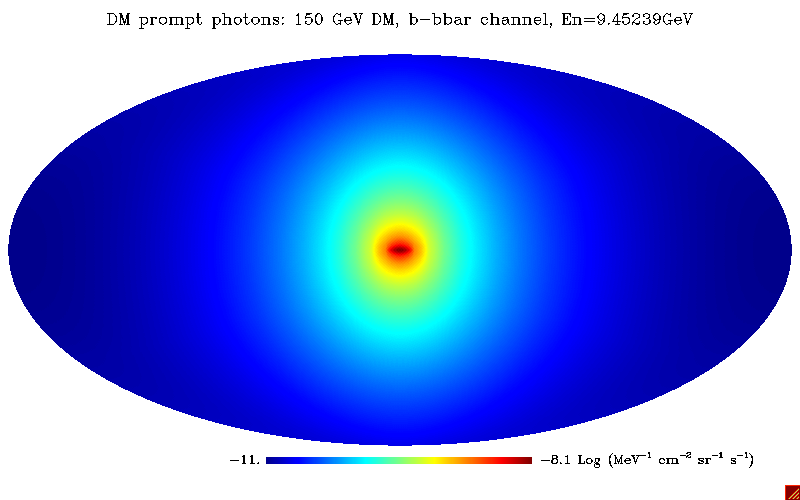}&
\includegraphics[width=0.45\columnwidth]{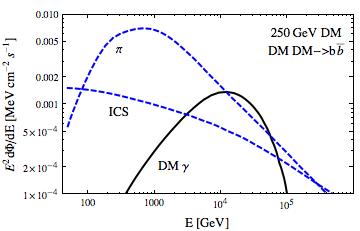}
\end{array}$
\end{center}
\vspace{-1.0cm}
\begin{center}$
\begin{array}{cc}
\includegraphics[width=0.45\columnwidth]{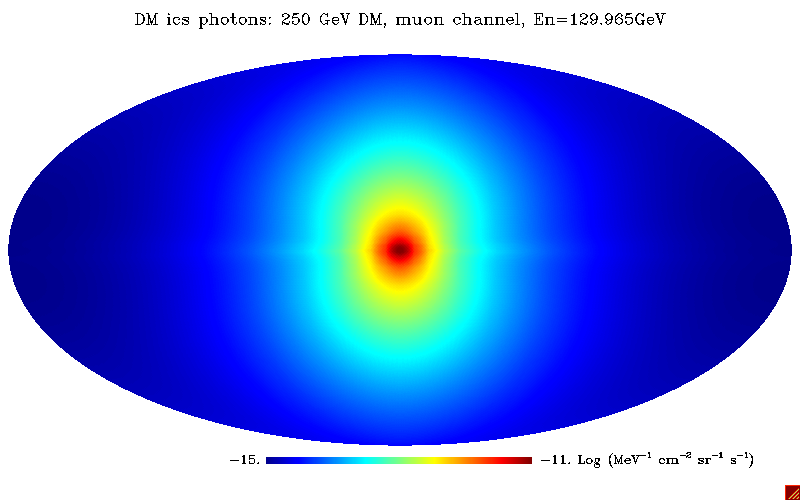}&
\includegraphics[width=0.45\columnwidth]{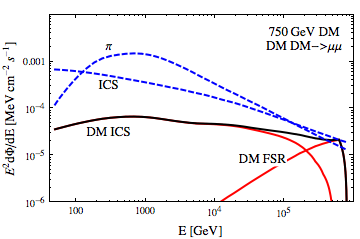}
\end{array}$
\end{center}
\vspace{-0.5cm}
\caption{Upper panel: All sky map (left) and the spectral shape of dark matter particles annihilation to $b{\bar b}$ (right). The typical spectra of the astrophysical contribution is shown for comparison. Lower panel: The same, for DM annihilation to $\mu ^+ \mu^-$. \label{dmmaps}}
\end{figure}

For the annihilation/decay spectra we considered two channels with distinctly different signatures: decay to $b{\bar b}$ channel, and to $\mu ^+ \mu^-$. The former case results in the spectrum which is similar for all channels in which DM annihilations/decays produce heavy quarks and gauge bosons (except for t quark), and is therefore predicted by most of the particle physics models. Gamma rays are then produced through hadronization and pion decay. The significant annihilation to leptonic channels, provided by the second scenario, is motivated by the PAMELA \cite{Adriani:2008zr} positron fraction and Fermi-LAT \cite{Abdo:2009zk} electron spectral measurements (see also \cite{Grasso:2009ma,Bergstrom:2009fa}). In this case, gamma rays are dominantly produced through radiative processes of produced electrons, as well as through the final state radiation, as shown in figure \ref{dmmaps}. DM maps are produced again with the \texttt{GALPROP} code, using, when propagation of electrons produced by DM is relevant, the same set of para
 meters of the astrophysical diffuse model. 

\section{Method}

To obtain the Dark matter limits shown here, we performed a binned likelihood fit with the \texttt{GARDIAN} tool,  using DataClean event class (where the charged particle contamination has been 
minimized, while maintaining $70\%$ of the acceptance of the Diffuse event class at high 
energies) and 21 months data, in the 800 MeV-100 GeV energy range. We further  mask the point sources from the 1FGL point sources catalogue. Notice, however, that we still keep the point sources model in the fit since, especially at low energy, some flux is leaking from the point sources outside the mask. Furthermore, we mask the bright Galactic plane, in order to minimize the uncertainty brought in for example, by the modeling of the gas maps and numerous point sources. Finally, we leave in the fit the following free parameters: overall normalization of H2, HI, IC and DM maps, normalization and spectral index of the isotropic component and residual contribution from point sources, for a total of 7 parameters.


\begin{figure}[t]
\begin{center}$
\begin{array}{cc}
\includegraphics[width=0.45\columnwidth]{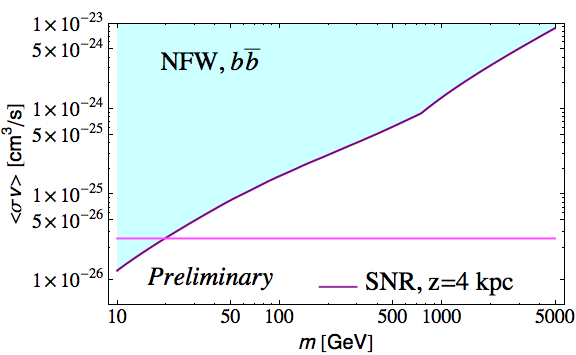}&
\includegraphics[width=0.45\columnwidth]{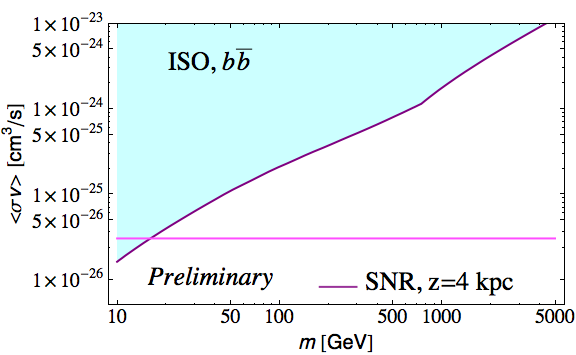}
\end{array}$
\end{center}
\vspace{-1.0cm}
\begin{center}$
\begin{array}{cc}
\includegraphics[width=0.45\columnwidth]{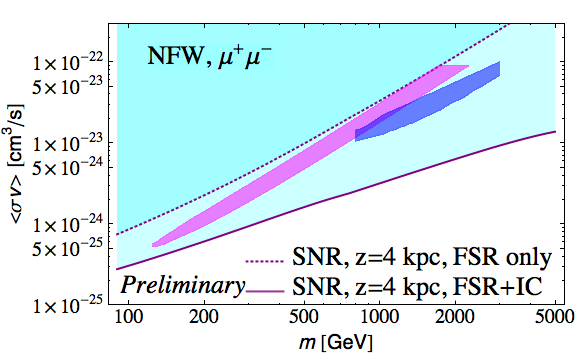}&
\includegraphics[width=0.45\columnwidth]{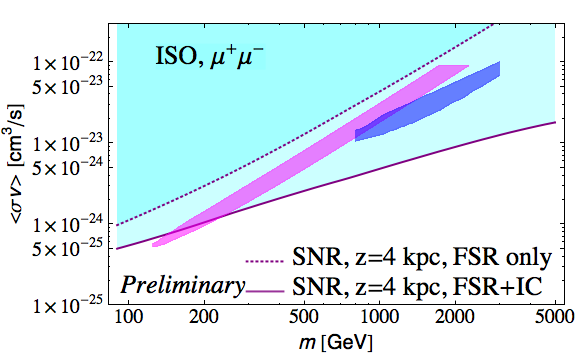}
\end{array}$
\end{center}
\vspace{-0.6cm}
\caption{DM limits, obtained by using the astrophysical model based on the cosmic ray source distribution  from direct observation of SNR, and a diffusive halo size of 4 kpc. Upper pannel: Limits on dark matter models in which DM annihilates to $b{\bar b}$, for DM distribution given by the NFW distribution (left) and isothermal distribution (right). Lower panel: The same, for DM annihilation to $\mu ^+ \mu^-$. In this case, regions of parameter space which provide a good fit to PAMELA \cite{Adriani:2008zr} (purple) and Fermi-LAT \cite{Abdo:2009zk} (blue) cosmic ray electron and positron data are shown, as derived in \cite{Bergstrom:2009fa}. \label{DMlimitsannihilation}}
\end{figure}

At the moment, a difficulty involved in the analysis is the fact that the fit always prefers a significant contribution of DM (for all DM channels and masses) making the procedure of setting an upper limit not well defined.  Conservatively, we use the procedure of attributing the observed excess to not well modeled background and we set upper limits at the 3 sigma confidence level above the best fit normalization of the DM contribution.  However, a further  drawback is that,  although the above astrophysical models derived for different  CR source distributions and different $z_h$  all describe reasonably well (not in a strict statistical sense, though) both the CR and the gamma-ray data  (see again  \cite{paper2}), the  DM limits depend significantly on the choice of the model. Again, here, we choose to be conservative and we quote as final upper limits the``worst"  among the ones obtained for the different models we explored, which come from a CR distribution modeled from the observed distribution of SNRs and a $z_h=$4 kpc.  

As shown on Figure \ref{dmmaps}, DM signal dominates in the Galactic Center and it has an extended latitude profile compared to the emission of astrophysical origin. 
Given these features, the first order effects on the DM limits are the behavior of the chosen CR source distribution in the vicinity of the GC (central few parsecs), which affects the emissivity in the region of the inner Galaxy, and the assumed size of the diffusive halo, which affects predicted latitude profile. We checked that the change in other propagation parameters (e.g. the index of the diffusion coefficient) have much smaller effect .

The cosmic ray source distribution is generally determined by the direct observation of supernovae, or its tracers (among which pulsars are prime candidates as they have a common predecessor), and it suffers from an observational bias, especially in the region of the inner Galaxy. There are currently only 46 observed SNR, and the distribution derived from their direct observation \cite{SNR}, in comparison to the one derived from the observation of pulsars \cite{Lorimer:2006qs,Yus}, predicts less sources in the inner galaxy, going to zero in the Galactic Center and with a weaker gradient. 
The difference is not large, however, the SNR source distribution is under-predicting the data in the inner Galaxy and the fit maximizes the likelihood by increasing the contribution of DM, which explains why the final DM limits are worse for this model, which we therefore conservatively chose.

Larger diffusive halo height fits the data better in terms of latitude profiles and the preferred halo size is found to be $\geq 10$ kpc. Smaller halo size typically under-predicts the observed latitude profile, and in that case too, the likelihood of the fit is increased with the addition of the DM component.

\begin{figure}[t] 
\begin{center}
\includegraphics[width=0.45\columnwidth]{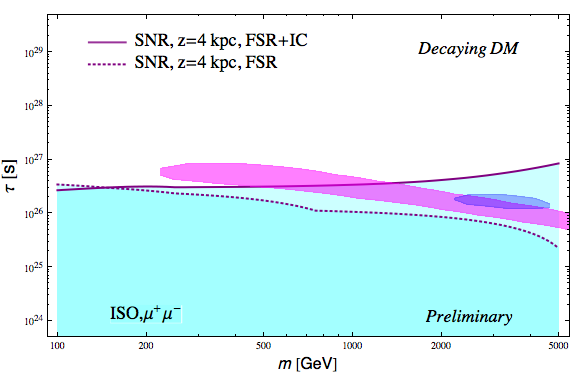}
\end{center}
\vspace{-0.6cm}
\caption{Limits on DM decay to $\mu ^+ \mu^-$, obtained by using the astrophysical model based on the cosmic ray source distribution  from direct observation of SNR, and a diffusive halo size of 4 kpc. In this case, regions of parameter space which provide a good fit to PAMELA \cite{Adriani:2008zr} (purple) and Fermi-LAT \cite{Abdo:2009zk} (blue) cosmic ray electron and positron data are shown, as derived in \cite{Papucci:2009gd}.\label{decay}}
\end{figure}

\section{Preliminary Results}

Overall, despite the various conservative choices, the limits are competitive with respect to the ones of other  analyses in gamma-rays like dwarfs or clusters \cite{Ackermann:2010rg, Abdo:2010ex}. In particular, as shown in Fig.\ref{DMlimitsannihilation} for masses around 20 GeV the thermal relic value of the annihilation cross section is reached, irrespectively of the DM Halo profile. More interestingly, considering annihilation into muons, the limits are significantly improved with respect to similar analyses performed without background modeling \cite{Papucci:2009gd,Cirelli:2009dv},  and   the DM interpretation of the PAMELA/Fermi CR features is ruled out, also irrespectively of the DM profile. 

We also consider constraints for the case of decaying DM which are shown in Fig.\ref{decay}. Also here the DM interpretation of the PAMELA/Fermi CR features is ruled out, apart a small region which is still allowed fitting Pamela only but not Fermi.

 

Further work is in progress along the lines outlined in this and \cite{ba} proceedings.


%
 


\end{document}